\documentclass[twocolumn]{aastex62}
\usepackage{graphicx}  
\usepackage{amsmath}
\usepackage{amssymb}   
\usepackage{xcolor}
\usepackage{nameref}
\usepackage{varioref}
\usepackage{hyperref}
\usepackage{cleveref}
\usepackage{ulem}
\usepackage[utf8]{inputenc}
\usepackage{graphicx}
\usepackage{delimset}
\usepackage{amssymb}
\usepackage{amsmath}
\usepackage{xcolor}
\usepackage{natbib}
\usepackage{ulem}

\newcommand{\chie}{\chi_{\rm eff}}
\newcommand{\WR}{*}
\newcommand{\dd}{\rm{d}}

\shorttitle{The Origin of Binary Black Holes Mergers}
\shortauthors{Piran \& Piran}
\begin{document}

\title[The Origin of Binary Black Holes Mergers]{The Origin of Binary Black Holes Mergers}
\author{Zoe Piran}
\affil{Racah Institute of Physics, The Hebrew University, Jerusalem 91904, Israel}
\author{Tsvi Piran}
\affil{Racah Institute of Physics, The Hebrew University, Jerusalem 91904, Israel}


\begin{abstract}
Recently \cite{IAS0} proposed a new pipeline to analyze LIGO-Virgo's  O1-O2 data and discovered eight new binary black hole (BBH) mergers, including  one with a  high effective spin, $\chie$. This discovery sheds new light on the origin of the observed BBHs and the  dynamical capture vs. field binaries debate. Using a tide-wind model, that characterizes the late phases  of binary evolution and captures the essence
of field binary spin evolution, we show that the observed $\chie$ distribution favors this model over capture.  However, given the current limited sample size, capture scenarios (isotropic models) cannot be ruled out. Observations of roughly a hundred merges will enable us to distinguish between the different formation scenarios. However, if as expected, both formation channels operate it may be difficult to resolve their exact fraction. 
\end{abstract}


\section{Introduction} 

The Ligo-Virgo Collaboration (LVC) discovery   \citep{LIGO_ANL, LIGO_cat} of merging binary black holes (BBH)  immediately posed a puzzle - what is the origin of these binaries? 
The numerous models that have been suggested   can be divided to two main groups:
``field evolution" models and dynamical capture models. In the former  the BBHs arose from binary massive stellar progenitor  \citep[e.g.][]{Phinney1991,Tutukov1993, belczynski2016Nature,mandel2016MNRAS,marchant2016,belczynski2017,stevenson2017NatCo,OShaughnessy17,belczynski2017,Qin2018,postnov2019,Bavera2019}. In the latter  each one  of the black holes formed on its own and the binary was assembled via a dynamical capture. These latter scenarios are further divided into two physically different subgroups. 
In the first, the BBHs are primordial \citep[e.g.][]{ioka1998PRD,bird2016PhRvL,sasaki2016PhRvL,blinnikov2016JCAP,kashlinsky2016ApJ}. In the second they formed from regular massive stars in various dense stellar environments in which the interaction with other stars led to the formation of the binary \citep{Sigurdsson1993,Portegies2000,Miller2009,OLeary2009,Kocsis2012,ISO,cluster_rod,cluster_Oleary,antonini2016ApJ,stone2017MNRAS,bartos2017ApJ,Fragione2018,Hoang2018,McKernan2018,Fragione2019,Secunda2019}. 

It has been long realized \citep{Mandel2010} that among the different parameters of a merging black hole binary, that can be easily recovered from the GW data\footnote{Spin measurements of individual BHs, that might be available in the future may shed a additional light on this question.},  the effective spin, $-1\le \chie  \le 1$ (the normalized component of the sum  of the two black holes spins projected  in the direction of the orbital spin), is the most informative parameter for studying the BBH origin \citep[see also][]{blinnikov2016JCAP,Kushnir,HotokezakaPiran2017ApJ}. The normalized spin of each black hole is defined as 
$\chi \equiv {c {\bf S}} \cdot \hat {{\bf  L} }/G m^2$, where  $m$ is  the black hole's mass, ${\bf S}$ its  spin vector, $\hat {\bf L}$ is the direction of the orbital spin and  $c$ and $G$ are the speed of light and Newton's constant. 
The binary's effective spin  is $\chie\equiv (\chi_{_{BH,1}}+q \chi_{_{BH,2}})/(1+q)$ with $q \equiv {m_2}/{m_1}$.

Capture scenarios don't provide a physical mechanism that links  the directions of spins of the individual black holes to the orbital angular momentum.  The former depends on the evolution of the individual black holes' progenitors  while the latter depends on their relative motion and the capture dynamics. 
This  results in what we denote as {\it``isotropic''} $\chie $ distribution  \citep{ISO}. In this case the spins are  randomly oriented relative to each other and to the orbital spin. Thus, we expect both positive and negative  $\chie$ values. As   
triple alignments are rare we don't expect large  $|\chie|$ values, and in any case  we expect equal number of positive and negative ones\footnote{Note that selection bias that depend on  $\chie$ \citep{campanelli2006PRL,roulet_mass} may lead to excess of positive $\chie$ events. But those can be taken into account only when sufficient data is available.}. In the following we will consider, following \citep{Farr}, three isotropic distributions, {\it low}, {\it flat} and {\it high},  
according to the distribution of the spins'  magnitudes (see Fig. \ref{fig:Isotropic_models}). 

On the other hand, various processes during the evolution of ``field binaries"  align  the stellar spins with the orbit. Among those are: (i) The binary formation process, if the progenitor stars formed from a rotating cloud; (ii)  Mass transfer  that   spins the recipient along the direction of the orbital motion; (iii) Tidal locking at various stages and especially   
at the late phases of the binary.
Some  mechanisms, in particular  winds,  reduce the progenitors spins. Others, such as kicks during the collapse \citep{tauris2017ApJ,mandel2016MNRASb,Wysocki2018}, {  randomize it, changing its direction and possibly magnitude\footnote{ A change in the magnitude may arise if the kick is not give at the CM of the collapsing star.}.} However,  no known mechanism  preferably rotates  the stellar spins into a   direction opposite to the orbital angular momentum. Hence 
we expect a positive correlation between individual spins and the orbit's angular momentum leading to a  preference of positive $\chie$ values over what is expected in an isotropic distribution (see Fig. \ref{fig:time_params}). 

Numerous attempts to model the expected distributions of BBH mergers parameters (masses, mass ratios, spins etc..) within the ``field binaries" scenario have been carried out using a detailed population synthesis approach \citep[see e.g.][]{belczynski2016Nature,belczynski2017,Wiktorowicz2019,Bavera2019}. These models follow all stages of stellar evolution from birth to death using the best current understanding of each phase  and construct expected distributions of all observed parameters. However, various unknown factors concerning critical phases during the binary evolution \citep[see e.g.][for a review concerning the common envelope phase]{ivanova2013A&ARv}  exist.    Instead, following \citep{Kushnir,HotokezakaPiran2017ApJ,KT_ANL}, we consider here a minimal model that identifies tidal locking and winds that operate at the latest stages of the stellar evolution as the dominant mechanisms  that determine the black hole's spins and the system's $\chie$. We take into account all earlier effects by varying the initial conditions of this final stage. The virtue of this model is its simplicity.  It includes only two free parameters which are sufficient to capture the essence of the observed $\chie$ distribution. 
{ The simplicity of the model implies that it cannot capture some of  the rich features of this last phase \citep[see][for a detailed discussion]{Qin2018,Bavera2019}. These are particularly important concerning the impact of winds that we discuss below. } However,
this approach is justified given the relatively small size of the current observed data-set that provides us  limited statistical knowledge about the  population. 

This  model (as we elaborate in  \S \ref{sec:models} and Appendix \ref{App_model}) considers only the last phase of the binary after one of the stars, the primary, has already collapsed to a black hole and it exerts a tidal force on its companion. At the same time strong winds reduce the companion's  spin. The competition between tides and winds determines the final progenitor's spin\footnote{As we explain later we expect that  natal kicks at the BH formation are unimportant.}. To account for the uncertainty in the earlier phases of the stellar evolution we consider two drastically different initial conditions at the beginning of this phase: The secondary star is either non-rotating or it is fully synchronized with the orbital motion. As for the primary, we consider it to  either  follow a similar evolution as the secondary (namely spins and tides) or to be randomly oriented. 

LVC's O1-O2 events distribution is  approximately symmetric around $0$ with a rather low $\chie$ values. Such a distribution favors a low isotropic model \citep{Farr,Wysocki2018}. However, this sample didn't provide enough events to rule out ``field binary" scenarios  \citep{KT_ANL,Wysocki2018}. The recent re-analysis of LVC's O1-O2 data recovered the ten binary black hole (BBH) mergers detected by LVC and revealed eight new events \citep{IAS0,IAS1,IAS2,IAS3}. We denote the combined set of observation as the LVC-IAS data-set.  Here we consider the question whether the newly identified mergers shed new light on the origin of these BBHs.
  
This work extends the analysis carried out  by \cite{KT_ANL} for the LVC data to the larger LVC-IAS data-set. In addition to  a modified ``field binary"  model, we use a novel way to account for  errors in the estimated $\chie$ values and we evaluate the quality of the fits using  the Anderson-Darling statistic. 
The LVC-IAS data-set is still rather small, thus  one cannot expect that it will conclusively rule out or confirm any one of the models. 
Furthermore, it is possible that  some BBH merges form via field evolution while others are captured. Hence
we also ask the question how many mergers should be detected in order to enable us to distinguish  between ``isotropic" models and ``field binary" models and how many will be needed to distinguish between a pure ``isotropic" or ``field binary" model and a mixed one. 

\begin{figure*}[htb!]
   \centering
    \includegraphics[scale = 0.6]{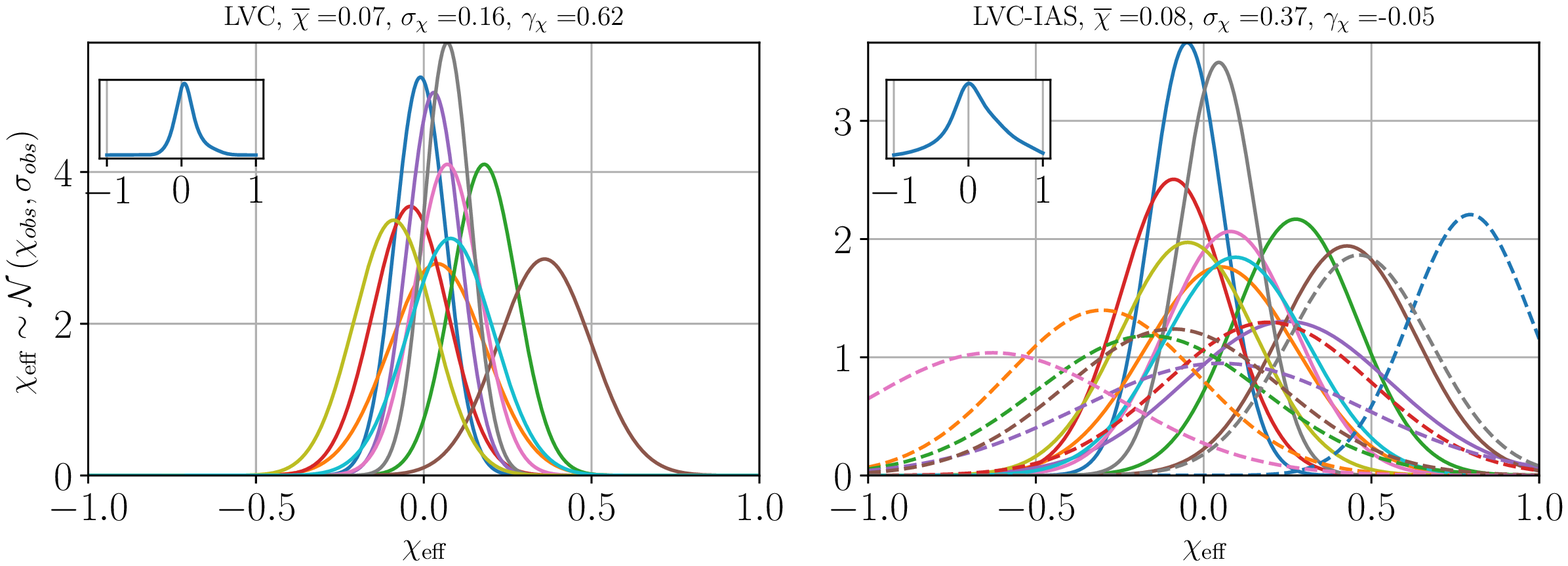}
    \caption{The distribution of the observed $\chie$ in the  LVC  (Left) and IAS (right) data. We have approximated each observation as a Gaussian whose mean value and $90\%$ credible interval are the values given in \citep{LIGO_cat} and \citep{IAS1} respectively. The inserts show the average distribution.  The title indicates  the mean $\overline{\chi}$, the standard deviation $\sigma_{\chi}$ and the skewness $\gamma_{\chi}$.}
    \label{fig:obs_p}
 \end{figure*}
 
 We describe the data-sets in \S \ref{sec:data}. We briefly discuss the models in \S \ref{sec:models} (leaving  some details  to a more technical Appendix \ref{App_model}) and describe the data analysis in \S \ref{sec:data.alanysis}.
The results are presented  in \ref{sec:results}, discussing first the results concerning the current data \S \ref{sec:results.current} and then in \S 
\ref{sec:results.future} we explore the question how many events are needed to distinguish between  different models. We summarize our findings in \S \ref{sec:conclusions}

\section{The Data} 
\label{sec:data}
The LVC analysis of the O1-O2 runs revealed ten BBH mergers \citep{LIGO_cat}. Recently \cite{IAS0} proposed  a novel pipeline for the analysis of  GW data. Estimated parameters of mergers  identified by both pipelines are within the errors of each other (see Table \ref{tab:data}  and  Fig. \ref{fig:obs_p}). However, the new  re-analysis of the O1 \citep{IAS1} and O2 data \citep{IAS2,IAS3} revealed  eight new BBH mergers.

In the following analysis we neglect possible mass/spin correlations. We evaluate the models over a fixed mass and compare them  to the unweighted\footnote{With respect to the mass.} observed distribution. This is natural in the isotropic scenario and valid for field  binary  scenarios if tidal locking and winds operate in the same manner across the progenitors mass range. Further, given the small size of the sample, such an assumption is essential.  For the same reason we neglect  \citep[see e.g.][]{Abbott2018c} bias that may arise from the  dependence of the GW horizon on the spin \citep{Campanelli2006a,roulet_mass}.

\section{The Models}
\label{sec:models}
\subsection{ Isotropic Models} 
\label{sec:isotropic} 
 The $\chie$ distribution is given by a weighted sum of two randomly oriented (isotropic) normalized spin vectors ${\bf s}_i$: 
\begin{equation}
\chi_{\rm{eff-iso}}=\frac{{\bf s}_1\cdot\hat{\bf L}+q{\bf s}_2\cdot\hat{\bf L}}{1+q} \ .
\end{equation}
Following \cite{Farr} we consider three distributions defined by the distribution of  $|{\bf s_i}|$:    {\it flat},   or  dominated by either {\it low} or  {\it  high} spins.  The probability for a given  ${\bf s}$ value is: 
\begin{equation} 
p(|{\bf s_i}|) = 
\begin{cases}
2(1-|{\bf s_i}|) & { \it low}; \\
1  & { \it flat}; \\
2|{\bf s_i}| & { \it high}.
\end{cases}
\end{equation} 
We use $q=1$ (varying $q$ has  a minor effect, see  Fig. \ref{fig_sup:mass_dep} in Appendix \ref{Additional Tests}). 

\subsection{Field Binaries:}
\label{sec:fieldbinaries}
Given the complexity of binary evolution { \citep[see e.g.][]{Qin2018,Bavera2019}} we consider here a minimal model \citep{Kushnir,HotokezakaPiran2017ApJ,KT_ANL} that captures the critical ingredients during  the last phase of the binary: the interplay between tidal locking, that increases and aligns the spin, and winds, that diminish it.  
{ We assume that the two processes (tidal locking and winds) are decoupled and  we neglect the possible interplay between the two due to the fact that  winds  increase  the orbital separation and this weakens the tidal force. }

To account for the uncertainty in  earlier phases of the evolution we  consider different initial conditions for the beginning of this last phase. 
We briefly outline here the essential ingredients, focusing in particular on revisions that we have introduced  to the model used by \cite{KT_ANL}). 
We consider Wolf-Rayet progenitors, as those are massive enough and have small enough radii allowing the binaries to merge within a Hubble time. However, the considerations are not limited to those  and would be relevant to  final stages of most field binaries, provided that their radii are small enough to fit within an orbit that can merge in a Hubble time. Numerical factors concerning the stellar model that we use in Eqs. \ref{eq:tc}-\ref{eq:chi_syn} below may be different in such cases but the basic result holds.

{\it Coalescence:} We assume that at the time that the second BH forms the orbit is circular \citep[see e.g.][]{HotokezakaPiran2017ApJ,prev_obs} with a radius $a$. The corresponding coalescence time is:
\begin{align}
t_c \approx 10\textrm{Gyr} ~  \brk*{\frac{2q^2}{1+q}} \brk*{\frac{a}{44 R_{\odot}}}^4 \brk*{\frac{m_2}{30 M_{\odot}}}^{-3} \ .
\label{eq:tc}
\end{align}
{ We use this equation  to express the orbital separation in terms of $t_c$. Consequently the $t_c$ distribution, discussed below, determines the orbital separation distribution and vice versa. }

{\it Synchronization:}  { The synchronization of the spin of a massive star due to the tidal force exerted by the companion  has been studied in different contexts by numerous authors \citep{Brown+2000,Izzard2004,Petrovic2005,Cantiello+2007,vandenHeuvel2007,detmers2008A&A,Eldridge2008}  and more recently  by \cite{Qin2018} within the context of BBH mergers. Here we characterize the effects by the time scale, $t_{syn}$, to synchronize the star spin with the orbit} \citep{Kushnir}:
\begin{align} 
\label{eq:tsyn}
t_{syn}\approx 10 ~\textrm{Myr} ~ q^{-{1}/{8}} \brk*{\frac{1+q}{2q}}^{{31}/{24}} \brk*{\frac{t_c}{1\textrm{Gyr}}}^{{17}/{8}} \ .
\end{align}
If fully synchronized with the orbit the  spin of the star is aligned and its normalized value is:
\begin{align} \label{eq:chi_syn}
\chi_{syn} \approx&~ 0.5~q^{{1}/{4}}& \brk*{\frac{1+q}{2}}^{{1}/{8}} \brk*{\frac{\epsilon}{0.075}} \brk*{\frac{R_2}{2R_{\odot}}}^{2}  \nonumber \\
&&\brk*{\frac{m_2}{30 M_{\odot}}}^{-{13}/{8}} \brk*{\frac{t_c}{1\textrm{Gyr}}}^{-{3}/{8}}\ ,
\end{align}
where $\epsilon \equiv {I_2}/{m_2R^2_2}$ relates the star's moment of inertia, $I_2$, to its mass and radius, $m_2$ and $R_2$.

{\it Winds:} Strong winds that operate at late phases of the stellar evolution lead to angular momentum loss, characterized by  $t_{w} \equiv 
\chi_{*} /\dot \chi_{*}$, where $\chi_*$ is the star's normalized aligned spin and $\dot \chi_*$ its angular momentum weighted\footnote{Wind from the equatorial plane  carries more angular momentum than the average specific angular momentum. If it dominates, angular momentum loss rate is  faster than mass loss rate.} loss rate. { As mentioned earlier, we neglect the winds' impact on the orbital separation and with this the coupling between winds and tidal locking
\citep{Qin2018}.  Additionally, the larger metallicity of BBHs  that form recently (and hence merge with smaller $t_c$ values reflecting smaller initial separations) increases the effects of winds relative to BBHs that formed earlier at larger initial separations. Both effects enhance the winds' impact. These effects will be incorporated within our model by shorter $t_w$ values that will indicate stronger winds. }

{\it Initial values:} We  consider initially  synchronized stars $ \chi_{*}(0)=\chi_{syn}$ or 
non-rotating stars $ \chi_{*}(0)=0 $, denoted by a subscript $_{syn,0}$ respectively.
These two extreme initial conditions reflect the large uncertainty in the earlier evolution of the stars.  

{\it  Evolution:} 
The combined effects of tidal forces and winds on the stellar spin yield \citep{Kushnir,HotokezakaPiran2017ApJ}:
\begin{equation} 
\label{eq:chi_star}
	\frac{\textrm{d} {\chi_{*}}}{\textrm{d}{t}} = \frac {({\chi_{syn}}-  \chi_{*})^{{8}/{3}}}{{t}_{syn}(t_c)}
	 - \frac{{\chi}_{*}}{{t_w}} \ . 
\end{equation}
We evolve $\chi_*$ over the lifetime of the  star $t_{\WR}$  to obtain the final spin $\chi_{*} (t_{\WR})$ (see Fig.\ref{fig:SM_evolution} in Appendix \ref{App_model}). 
We fix the stellar life time at  $t_{\WR}=0.3\textrm{Myr}$. Changing this value will merely amount to re-scaling the other time scales $t_{syn}$ and $t_w$. 
The  ratio $ \chi_{*}(t_{\WR})/{\chi_{syn}}$  depends on  ${t_{w}}/{t_{syn}(t_c)}$ and 
${t_{w}}/{t_{\WR}}$. While the latter is of order unity, the former varies over a large range, due to the strong dependence of $t_{syn}$ on $t_c$ (see Eq. \ref{eq:tsyn}).

{\it  Collapse:} If $\chi_{*} (t_{\WR})\leq 1$,  the entire star implodes to a BH  with  $\chi_{_{BH}} = \chi_{*} (t_{\WR})$. If  $\chi_{*} (t_{\WR})>1$, a fraction of the matter must be ejected carrying the excess angular momentum and  $\chi_{_{BH}} \lesssim 1$ \citep{stark1985PhRvL}. 
Observations of massive ($> 10 M_{\odot}$)  Galactic BHs X-ray binaries indicate that massive BHs form in situ in a direct implosion and without a kick \citep{prev_obs}. Therefore, we disregard here possible natal kicks \citep[see e.g.][]{tauris2017ApJ,mandel2016MNRASb,Wysocki2018} that may tilt the spin and randomize it.

{\it  Single/double synchronization:}  
In the Single Aligned ($\textrm{SA}$) scenario  tidal locking and winds operate  only on the secondary (the lighter) star and the resulting effective spin, $\chi_{_{BH,2}}$, is calculated as outlined above (see Appendix \ref{App_model} for details).  We take $\chi_{_{BH,1}}$ to be distributed as {\it flat} isotropic\footnote{\citep{KT_ANL} 
assumed in this scenario that the primary always has  $\chi_{_{BH,1}}=0$.}.
We also consider a Double Aligned ($\textrm{DA}$) scenario  in which tidal locking and winds operate on both stars in a similar manner.

{\it  Rates and delay distribution:}  
We assume that the BBHs formation rate follows the star formation rate (SFR)
\citep{SFR}:
$R_{SFR}\brk*{z}\propto{\brk*{1+z}^{2.7}}/\{{1+[({1+z})/{2.9}]^{5.6}}\}$.  
This is uncertain as the progenitors are very massive stars, but we have verified (see Appendix \ref{Additional Tests} Fig. \ref{fig_sup:rate}) that our predictions don't depend strongly on the details of the BBH formation rate. { In particular we also consider a formation rate that follows LGRBs that in turn follow a low metallicity population. }

The mergers' rate follows the formation rate with a time delay $t_c$ whose probability is assumed to be distributed as $ p_{\textrm{obs}}\brk*{t_c} \propto t_c^{-1}$  for $t_c > t_{c,min}$.  { This last parameter, $t_{c,min}$ is one of the critical parameters  of the model as $t_c$ determines the  separation between the two progenitors just before the second collapse. $t_{c,min}$ corresponds, therefore to the minimal separation. The separations above this minimal one are equally distributed in the logarithm.  }

A detailed description of the implementation of the model and the calculation of the resulting probability distribution is given in Appendix \ref{App_model}. 

\section{Data Analysis:} 
\label{sec:data.alanysis}
 To estimate the validity of each model we use the \cite{anderson1952} test. 
The Anderson-Darling (AD) statistic is model dependent. To allow for a proper comparison we obtain the significance level, given in Table \ref{tab:AD_test}, of each model independently. 
For a given model, described by a distribution $p(\chie)$, the significance test is performed as follows: $(i)$ We sample $\mathcal{N}= 18$ noiseless data points from $p(\chie)$. $(ii)$ We add an error sampled from a centered Gaussian with a standard deviation, $\overline{\sigma}_{\chie}=0.15$ (the average standard deviation in the observed $\chie$ estimates, see  Table \ref{tab:data}) to each data point. $(iii)$  We evaluate the $A^{2}$ statistic of the obtained data-set. 
$(iv)$ Repeating this process $10^6$ times gives an empirical distribution of ${A^2}$ from which we obtain the acceptance values (see
Table \ref{tab:AD_test}).

\begin{table*}[htb!]
\begin{center}
\begin{tabular}{||c || c | c | c | c | c | c | c | c | c | c |c |c |c|c|c||} 
\hline \hline
  & $99\%$ & $90\%$ & $80\%$ & $70\%$ & $60\%$ & $50\%$ & $40\%$ & $30\%$ &  $20\%$ & $10\%$ & $5\%$ & $4\%$& $3\%$& $2\%$& $1\%$ \\ [0.5ex] 
\hline \hline
$\rm{SA}_{0}, \rm{SA}_{syn}$ &$0.18$ &$0.33$  & $0.44$ & $0.55$ & $0.66$ & $0.8$ &$0.96$ & 
$1.18$ & $1.49$ & $2.06$  & $2.69$ & $2.88$ & $3.15$ & $3.53$ & $4.19$ \\ [1ex]
\hline
$\rm{DA}_{0},\rm{DA}_{syn}$ &$0.18$ &$0.33$  & $0.45$ & $0.55$ & $0.67$ & $0.8$ &$0.96$ & 
$1.18$ & $1.5$ & $2.09$  & $2.71$ & $2.91$ & $3.18$ & $3.55$ & $4.24$  \\ [1ex]
\hline
$(\rm{SA}_{0,syn} + \rm{DA}_{0,syn})/2$ &$0.18$ &$0.33$  & $0.44$ & $0.55$ & $0.67$ & $0.8$ &$0.96$ & 
$1.18$ & $1.5$ & $2.08$  & $2.7$ & $2.9$ & $3.17$ & $3.55$ & $4.16$  \\ [1ex]
\hline
$\rm{ISO}_{\rm{low}}, \rm{ISO}_{\rm{flat}},\rm{ISO}_{\rm{high}}$  &$0.18$ &$0.33$  & $0.44$ & $0.55$ & $0.66$ & $0.79$ & $0.95$ & 
$1.16$ & $1.47$ & $2.05$  & $2.64$ & $2.85$ & $3.11$ & $3.49$ & $4.12$ \\ [1ex]
\hline \hline
\end{tabular}
\caption{Acceptance values of Anderson-Darling test statistic $A^2$ for the different models. }
\label{tab:AD_test}
\end{center}
\end{table*}

To test how many events are required to distinguish between two models we carry out the following procedure. We 
choose one distributions, denoted $p_{ref}(\chie)$, as  describing the ``real world" and compare it to a test distribution, denoted  $p_{cmp}(\chie)$. To do so we obtain a data-set by sampling $\mathcal{N}$ events from $p_{ref}(\chie)$. We consider those as our ``observed" events and we carry out the same analysis as described earlier to test the compared model against this data-set using the AD test. We perform this  over a range  of sample sizes, $\mathcal{N}$. 

Before  comparing any model distribution to the data we must take into account the errors in the estimated $\chie$ values.
To do so, for each model  described by a parameter set $\lambda$,  we evaluate the theoretical probability,  $p_{\rm{th}}(\chie; \lambda)$  (see Appendix \ref{App_model} for details). We then account for the errors by convolving $p_{\rm{th}}(\chie; \lambda)$  with a Gaussian characterized by  $\bar \sigma_{\chie}=0.15$. The final model prediction is given by:  
 \begin{equation}
 p(\chie;\lambda) = \int_{-1}^{1} p_{\rm{th}}(\chie';\lambda)\frac{e^{- \brk*{\chie-\chie'}^2/2 \bar \sigma_{\chie}^2}}{\sqrt{2 \pi \bar \sigma_{\chie}}} \textrm{d}\chie' \ . 
 \end{equation}

{ As  some of the  events have a lower $p_{astro}$ values  we have also analyzed the data taking this probability into account.  Each one of the events was given  a weight that is proportional to its $p_{astro}$. This has a minimal effect on the results (see Fig. \ref{fig:Isotropic_models} in Appendix \ref{Additional Tests}). }

\section{ Results} 
\label{sec:results}
We  compare  the current data to different models in \S \ref{sec:results.current}. Given the best model in each category  we then address in \ref{sec:results.future} the question:  how many events are required to obtain a statistically significant result that will distinguish between the two categories? 

\subsection{Current Data}
\label{sec:results.current} 
\begin{figure}[htb!]
   \centering
    \includegraphics[scale = 0.87]{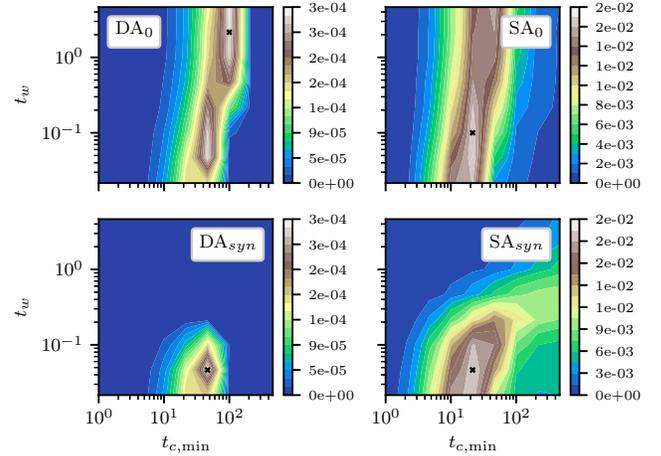}
    \caption{The likelihood $\mathcal{L}$ of the four different field binary models  over a range of time-parameters, $t_{w} \in \brk[s]*{0.03, 5}\textrm{Myr}$ and $t_{c,\min} \in \brk[s]*{1, 1000}\textrm{Myr}$.  The Maximum-Likelihood values are marked with a $*$.}
    \label{fig:mle}
 \end{figure}

We compare the observed LVC-IAS $\chie$ distribution to the expected ones for three isotropic distributions: {\it low, flat} and {\it high}, \citep[as defined in][]{Farr} and the four field binary models: $\textrm{SA}_{0,syn}$ and $\textrm{DA}_{0,syn}$, described above. We optimize the parameters of the field binary models by performing a Maximum-Likelihood test (see Fig. \ref{fig:mle}). 

\begin{figure}[htb!]
   \centering
    \includegraphics[scale = 0.8]{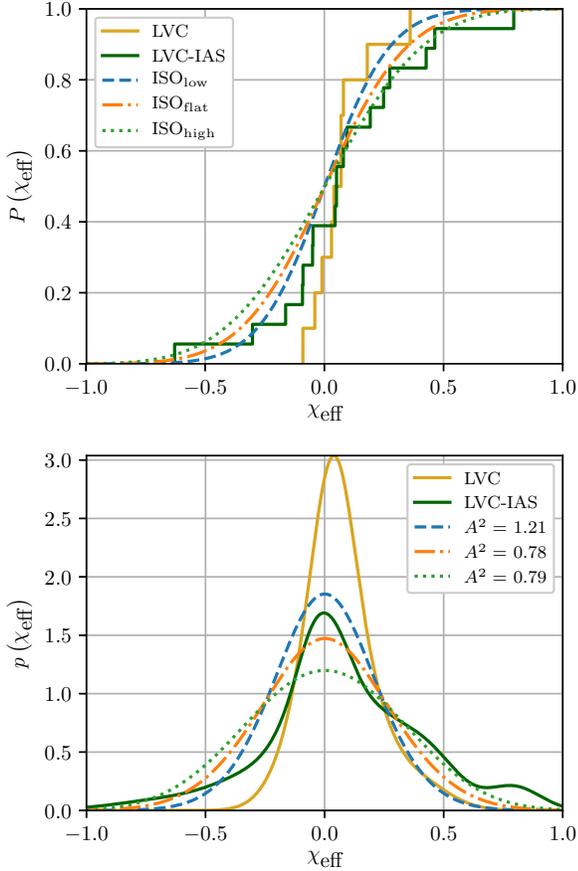}
    \caption{A comparison of the cumulative distribution, $P(\chie)$, (top) and the probability density function, $p(\chie)$, (bottom) of the {\it low}, {\it  flat} and {\it high} isotropic models (with observation errors added) with the LVC-IAS data. The AD statistic, $A^2$, is marked for each model. }
    \label{fig:Isotropic_models}
 \end{figure}

{\it Isotropic Models:} Fig. \ref{fig:Isotropic_models} depicts a comparison of the distributions of the three isotropic models to the LVC-IAS data.  All three isotropic models are acceptable. However,  the {\it high } model is favored whereas  the {\it low} model  was the most favorable  with the LVC data \citep{Farr}.

\begin{figure}[htb!]  
\centering 
\includegraphics[scale = 0.8]{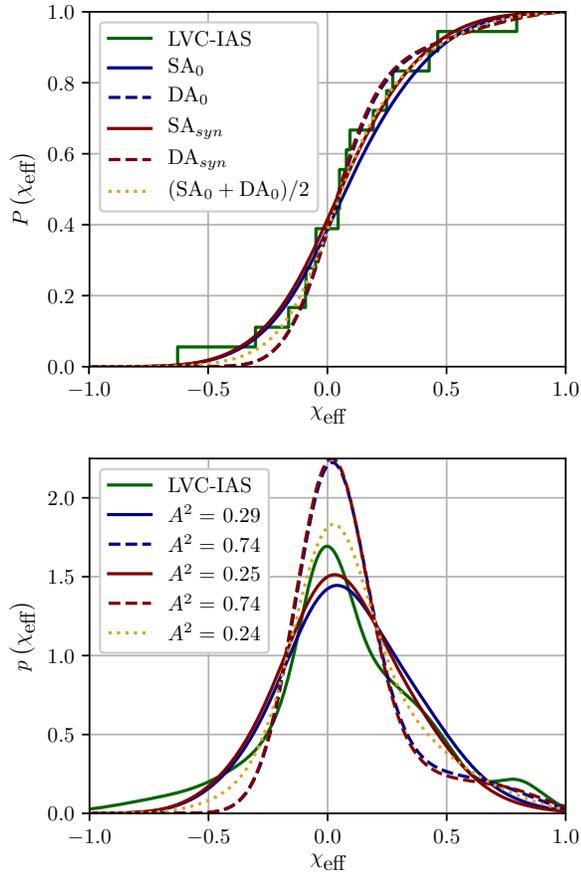} 
\caption{The cumulative distribution, $P(\chie)$, (top) and the probability density function, $p(\chie)$, (bottom) for the LVC-IAS data and different field binary models (using the best fitted parameters  $(t_{c,\min}, t_{w}) \textrm{Myr}$: ${\rm{SA}}_0: (20, 0.1)$, ${\rm{DA}}_0: (100, 1)$, ${\rm{SA}}_{syn}: (20, 0.05)$, ${\rm{DA}}_{syn}: (50, 0.05)$, and a mixed model,  $({\rm{SA}}_0 +{\rm{DA}}_0 )/2$, taken with $\rm{SA}_0$ parameters).
Note the excess of intermediate and high positive $\chie$  in these models.}    
\label{fig:time_params} 
\end{figure}

{\it Field Binaries:} 
The models depend on 
three time-parameters, $t_{\WR}, t_{c,\min}$ and $t_{w}$. We take $t_{\WR} = 0.3\textrm{Myr}$ as the typical\footnote{Variation of $t_{\WR}$ will amount to scaling of the two other time scales (see Appendix \ref{App_model} Fig. \ref{fig:SM_evolution}).}  lifetime
and use  Maximum-Likelihood  (see Fig. \ref{fig:mle}) to determine the best  $t_{c,\min}, t_{w}$ values.   
We find good fits (see Fig. \ref{fig:time_params}) for all models. The two {\rm SA} models, initially unsynchronized  and synchronized, result in almost identical distributions (using different  parameters). Similarly, the two {\rm DA} models give identical distributions. 

$\textrm{SA}_0$  stands out as the preferred model with the highest Maximum-Likelihood and the most reasonable physical parameters \citep[see][]{KT_ANL}: $t_{c,\min} = \brk[c]*{10 - 100 ~\textrm{Myr}}$ (corresponding, for $m_i\approx30M_{\odot},\ q=1$,  to  $a=4- 7\cdot 10^{11}~ \textrm{cm}$) and $t_w=0.1-5~\textrm{Myr}$, reflecting a wide range of wind time scales. { These values, that are on the lower side and correspond to a rather strong winds reflect probably the fact that our model underestimates somewhat the effect of  winds.} 
 The Maximum-Likelihood of $\textrm{SA}_{syn}$ is comparable to the one of $\textrm{SA}_0$ but the former requires somewhat stronger winds ($t_w < 0.1\textrm{Myr}$) and is valid at a more confined range. 
$\textrm{SA}_0$ and $\textrm{DA}_0$ have a comparable broad range of allowed physically acceptable parameters but the latter has a smaller maximal likelihood. The $\textrm{DA}_{syn}$ model has the smallest feasible parameter phase space and seems least likely.  We also consider, as an example, a model that  combines the two with  $0.5 (\textrm{SA}_0+ \textrm{DA}_0$) using the best fit parameters of the $\textrm{SA}_0$ model.  Even without optimizing the relative ratio of the two cases and the model parameters, this model fits the data slightly better than all others. 
When considering different stellar models the  numerical factors that appear in Eqs. \ref{eq:tsyn},\ref{eq:chi_syn} as well as the typical stellar life time, $t_{\WR}$, vary.  However variations in these factors will only amount to a variations in the best fit parameters  and not to the quality or the overall behavior of the different scenarios. 

\begin{figure}[htb!]  
\centering    
\includegraphics[scale = .8]{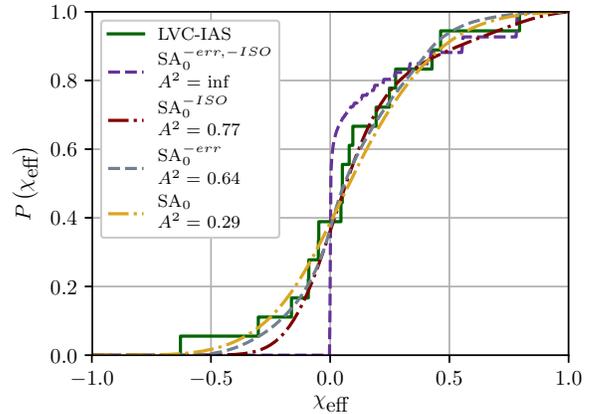}    
\caption{Same as Fig. \ref{fig:time_params} comparing between different $\textrm{SA}_0$ models. 
The curve 
{-err,-ISO} corresponds to one of the models in \cite{KT_ANL}  that arise when measurements errors and an isotropic spin component for the primary BH are not taken into account. Clearly such model has no negative $\chie$ values. 
}     
\label{fig:err_cmp} 
\end{figure}
Fig. \ref{fig:err_cmp} depicts different $\textrm{SA}_0$ models demonstrating the effect of the errors on the model   as well as the contribution of the addition of an isotropic spin $\chi_{1}$ into the $\textrm{SA}$ scenario. Both influence the resulting $\chie$ distributions giving a non-zero probability to $\chie > 0.5$ and to $\chie < 0$ events\footnote{Natal kicks, that we have 
neglected, can also give rise to negative $\chie$ values \cite{Wysocki2018}.}.

\subsection{Future Estimates}
\label{sec:results.future} 
As stated earlier, the current data-set is insufficient.  Even the least preferred  model, the low-isotropic, is consistent at $\sim 20\%$ with the data. 
We turn now to address the following question:  Assuming that one of the models is the correct one how many mergers are needed to rule out the others?  To do so we choose one of the models as the fiducial one, characterized by a distribution $p_{ref}(\chie)$. 
We can now test any  model, denoting it's probability density as $p_{cmp}(\chie)$, against the reference model. To do so we carry out the following procedure: (i) For each sample size, $\mathcal{N}$, We create an AD acceptance  table for $p_{cmp}(\chie)$.  (ii) We sample $10^5$ different data-sets (numerical tests reveal that this number is sufficient) of size $\mathcal{N}$ from $p_{ref}(\chie)$. (iii) We compute the average acceptance percentage of these data-sets. We repeat this procedure for $\mathcal{N}$ values ranging from $20$ to a few hundred choosing different models as the fiducial one and as the tested ones. 

Within the ``field evolution"  models we consider the ${\rm SA}_0$ model and the ${\rm DA}_0$ with the best fit parameters over the current data-set. We compare those to the three isotropic models, {\it low,} {\it flat} and {\it high}. We also consider a mixed model in which  $50\%$ of the event are ${\rm SA}_0$ field evolution binaries while the other $50\%$  are flat isotropic. 

\begin{figure}[htb!]  
\centering 
\begin{tabular}{@{}c@{}}
\includegraphics[scale = 0.8]{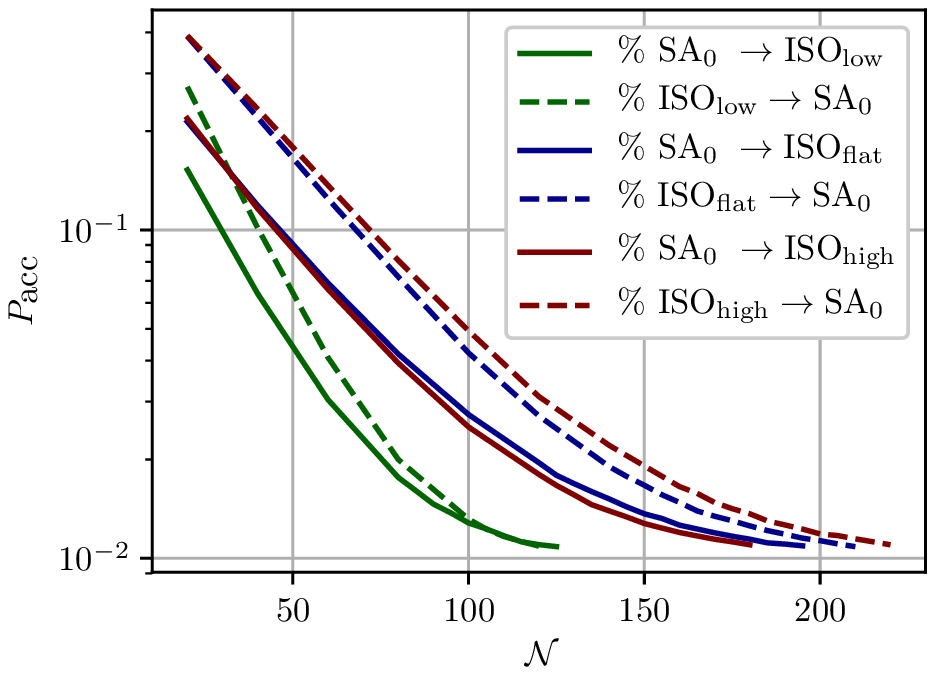}  \\ 
\includegraphics[scale = 0.8]{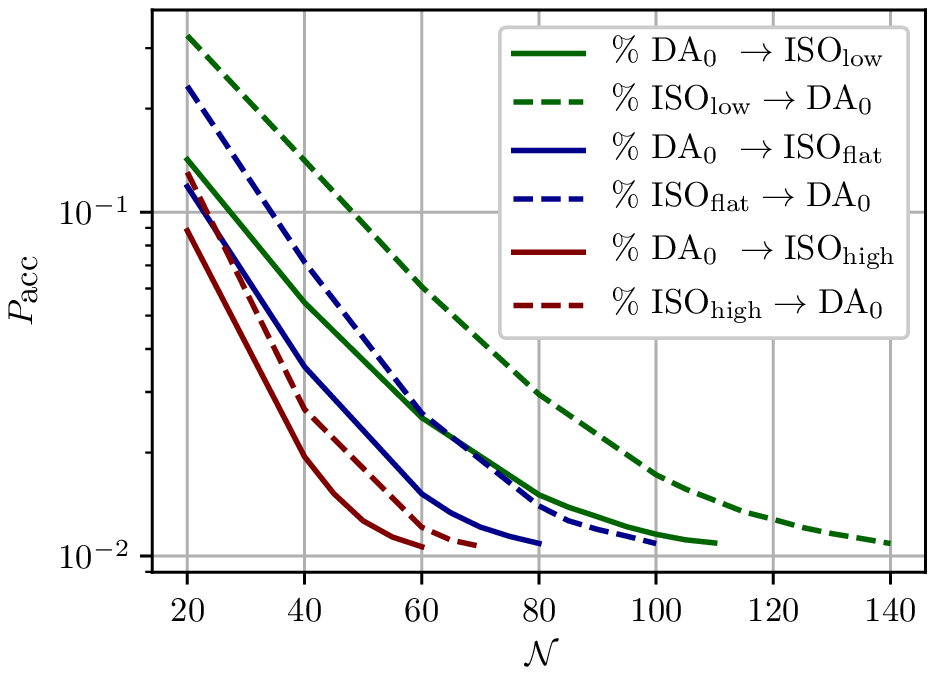} \\ 
\end{tabular}
\caption{Probability of acceptance, $P_{acc}$, (or  rejection that is $1-P_{acc}$) level of different models as a function of the number $\mathcal{N}$ of events detected. We compare the field evolution models, $\rm{SA}_0$ (top) and $\rm{DA}_0$ (bottom), with the three isotropic models low flat and high. Solid curves assume that the field evolution model describes the ''real world" and depict the acceptance significance of the isotropic models. Dashed curves assume that the isotropic models are the correct one and estimate the acceptance level for the respective field model. As expected it is easier to distinguish between the low isotropic and the $\rm{SA}_0$ model than to distinguish between the flat and high isotropic models and $\rm{SA}_0$. Further, a fewer events are needed (as compared to $\rm{SA}_0$) to distinguish between the $\rm{DA}_0$ field evolution model and the isotropic models}  
\label{fig:compare} 
\end{figure}

Fig. \ref{fig:compare} (top) depicts the resulting acceptance (1-rejection) probability of different tested models. It appears that $50-100$ ($150-250$) mergers are require to distinguish  the ${\rm SA}_0$ model from the isotropic ones at the  $5\%$ ($1\%$) confidence level.  The ${\rm DA}_0$ model includes more positive high spin events and fewer negative spin ones. Hence, as expected, it is easier to distinguish it from the isotropic models. As shown in  Fig. \ref{fig:compare} (bottom) $30-60$ ($50-120$) mergers are sufficient to distinguish between the ${\rm DA}_0$ model and the different isotropic models at the $5\%$ ($1\%$) confidence level. 
A caveat in the above estimate is that it cannot account for possible variations in the best fit parameters of the field binaries models ${\rm SA}_0$, ${\rm DA}_0$ that may arise in a large data-set.
The situation is more complicated when we consider mixed models that combine both field binaries and capture. A few hundred mergers are needed to distinguish between these models and ``pure" field binaries or ``pure" capture models. 
 
\begin{figure}[htb!]  
\centering 
\includegraphics[scale = 0.8]{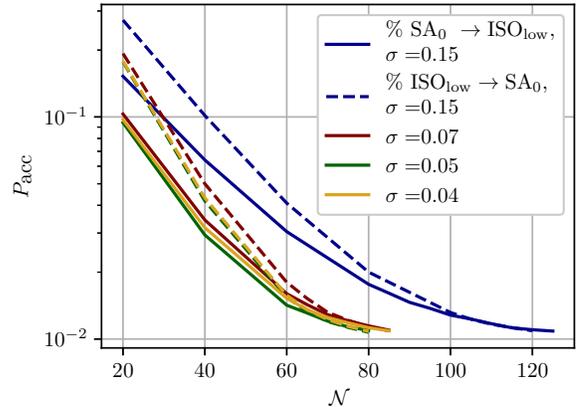} 
\caption{The acceptance probability as a function of the number of mergers (same as Fig. \ref{fig:compare}) between the $\rm{SA}_0$ and the {\it low} isotropic models. Solid lines test the acceptance of a low isotropic model assuming that the correct model is $\rm{SA}_0$.  Dashed lines consider the opposite.   The  standard deviation in measuring $\chie$ varies from 0.15 as in the current observed sample to  0.04. }  
\label{fig:compare2} 
\end{figure}

One may wonder whether a few strong events whose $\chie$ can be determined at a higher accuracy  
can change these conclusions. To check this we carried out  the same test using now  different values of $\bar \sigma_{\chie}$, the  standard deviation in the estimation of  $\chie$ (see Fig. \ref{fig:compare2}).  As expected  fewer events would suffice with a lower $\bar \sigma_{\chie}$. If $\bar \sigma_{\chie}$ is a quarter of its current value $25-50$ ($80-120$) events are needed to  distinguish at the $5\%$ ($1\%$) confidence level between the ${\rm SA}_0$ model and the isotropic ones. 

\section{Conclusions}
\label{sec:conclusions}
The observed low effective spins, that were centered around $0$, in the LVC O1-O2 sample favored low spin isotropic distributions \citep{Farr} and hence capture scenarios. We have shown here that while the combined LVC-IAS data-set 
that includes a high $\chie$ binary, cannot rule out any model it favors field binaries over capture. 

Within the field binary models the high $\chie$ merger implies a significant fraction of short ($t_c \sim 20\textrm{Myr}$) mergers, namely BBHs that at formation had small,  but reasonable {($4-7 \times 10^{11}\textrm{cm}$}), separations. 
Overall the LVC-IAS sample brackets nicely the phase space of the field binary model with $10\textrm{Myr}\lesssim t_{c,min} \lesssim 100\textrm{Myr}$, $0.05\textrm{Myr}\lesssim t_w \lesssim 5\textrm{Myr}$.

While the isotropic scenario is disfavored
it is not ruled out. Among those models the {\it high} variant becomes the most favorable and the {\it low} the least.   It is interesting to note that recently \cite{Romero-Shaw19} have shown that the eccentricity of all the events in the LVC sample are smaller than $0.02$ to $0.05$, whereas a capture scenario suggests that $5\%$ of the events should have larger  eccentricity.   Again, while this result doesn't rule out the capture scenarios they support our findings.  
Clearly, a mixture of field binaries and capture is possible. In this case we expect that the former will be dominant. However, given the limited data we didn't explore this possibility here.  Considering future observations we note that the hallmark of the field binaries scenario is a preferably positive $\chie$ distribution with a few large positive $\chie$ mergers. At the same time, unless kicks are very significant and dominate the BHs spin distribution, large negative $\chie$ will pose a problem for the field binary model.   
We have  shown that for the models considered here we will need $30-250$ events, depending on the details of the model and the level of confidence required to distinguish between the two scenarios.  Higher S/N  data  that has a better determined $\chie$ value would require a fewer events.
Hundreds of events will be needed to determine the ratio of capture to field evolution events in  mixed model that includes both capture and field binaries, or
to distinguish those from pure capture or pure field evolution models.

\section*{Acknowledgements}
We thank Tejaswi Venumadhav, Barak Zackay, Javier Roulet, Liang Dai  and  Matias Zaldarriaga  for sharing their  data with us prior to publication and we  acknowledge fruitful discussions with 
Ofek Birnholtz, Giacomo Fragione, Kenta Hotokezaka, Ehud Nakar, Bill Press, Nicholas C. Stone, and Barak Zackay.  
The research was supported by an advanced ERC grant (TReX), by the I-Core center of excellence of the CHE-ISF (TP) and by the Israeli Council for Higher Education (ZP).


\setcounter{figure}{0}
\setcounter{table}{0}
        \renewcommand{\thefigure}{A\arabic{figure}}%
        \renewcommand{\thetable}{A\arabic{table}}%
\appendix
\section{Data}
\begin{table*}[htb!]
\begin{center}
\begin{tabular}{||c | c c | c c | c c | c ||} 
\hline \hline
Event & \multicolumn{2}{| c |}{${m_1}/{M_{\odot}}$ } & \multicolumn{2}{| c |}{${m_2}/{M_{\odot}}$} &
 \multicolumn{2}{| c |}{$\chie$} & $p_{\mathrm{astro}}$ \\
\hline
 & LVC & IAS & LVC & IAS & LVC & IAS & IAS\\ [0.5ex] 
\hline\hline
 GW150914 & $35.6^{+4.8}_{-3.0}$ & $35.9^{+4.47}_{-4.45}$ & $30.6^{+3.0}_{-4.4}$ & $29.6^{+3.5}_{-3.6}$ & $-0.01^{+0.12}_{-0.13}$ &$-0.05^{+0.11}_{-0.11}$ & $*$ \\ [1ex]
\hline
 GW151012 & $23.3^{+14.0}_{-5.5}$ & $27.3^{+12.03}_{-11.89}$ & $13.6^{+4.1}_{-4.8}$ & $12.11^{+4.25}_{-4.2}$ & $0.04^{+0.28}_{-0.19}$ & $0.05^{+0.23}_{-0.23}$ & $>0.99$ \\ [1ex]
\hline
 GW151226 & $13.7^{+8.8}_{-3.2}$ & $16.4^{+7.53}_{-7.5}$ & $7.7^{+2.2}_{-2.6}$ & $7.53^{+2.45}_{-2.47}$ & $0.18^{+0.20}_{-0.12}$ & $0.27^{+0.19}_{-0.19}$ & $*$ \\ [1ex]
\hline
 GW170104 & $31.0^{+7.2}_{-5.6}$ & $30.17^{+5.95}_{-5.96}$ & $20.1^{+4.9}_{-4.5}$ & $19.6^{+3.89}_{-3.85}$ & $-0.04^{+0.17}_{-0.20}$ & $-0.09^{+0.16}_{-0.16}$ & $>0.99$ \\ [1ex]
\hline 
GW170608 & $10.9^{+5.3}_{-1.7}$ & $17.4^{+11.43}_{-11.45}$ & $7.6^{+1.3}_{-2.1}$ & $5.8^{+2.66}_{-2.66}$ & $0.03^{+0.19}_{-0.07}$ & $0.25^{+0.3}_{-0.31}$ & $>0.99$ \\ [1ex] 
\hline
GW170729 & $50.6^{+16.6}_{-10.2}$ & $50.3^{+12.42}_{-11.5}$  & $34.3^{+9.1}_{-10.1}$ & $34.6^{+9.05}_{-9.03}$ & $0.36^{+0.21}_{-0.25}$ & $0.43^{+0.21}_{-0.21}$ & $>0.99$ \\ [1ex]
\hline
GW170809 & $35.2^{+8.3}_{-6.0}$ &  $36.1^{+7.8}_{-7.76}$ & $23.8^{+5.2}_{-5.1}$ & $23.6^{+4.95}_{-4.94}$ & $0.07^{+0.16}_{-0.16}$ & $0.08^{+0.19}_{-0.19}$ & $>0.99$ \\ [1ex]
\hline
GW170814 & $30.7^{+5.7}_{-3.0}$ & $31.0^{+4.53}_{-4.53}$  & $25.3^{+2.9}_{-4.1}$ & $24.98^{+3.29}_{-3.29}$ & $0.07^{+0.12}_{-0.11}$ & $0.05^{+0.11}_{-0.11}$ & $>0.99$ \\ [1ex]
\hline
GW170818 & $35.5^{+7.5}_{-4.7}$ & $35.4^{+5.92}_{-5.91}$ & $26.8^{+4.3}_{-5.2}$ & $26.87^{+4.59}_{-4.6}$ & $-0.09^{+0.18}_{-0.21}$ & $0.05^{+0.2}_{-0.2}$ & $>0.99$ \\ [1ex]
\hline
GW170823 & $39.6^{+10.0}_{-6.6}$ & $39.5^{+7.34}_{-7.43}$  & $29.4^{+6.3}_{-7.1}$ & $28.5^{+5.91}_{-5.91}$ & $0.08^{+0.20}_{-0.22}$ & $0.09^{+0.22}_{-0.22}$ & $>0.99$ \\ [1ex] 
\hline
GW170121 & $-$ & $31.8^{+6.56}_{-6.58}$ & $-$ & $23.9^{+5.07}_{-5.05}$ &  $-$ & $-0.3^{+0.29}_{-0.29}$ & $>0.99$ \\ [1ex]
\hline
GW170727 & $-$ & $40.1^{+8.34}_{-8.35}$ & $-$ & $29.1^{+6.6}_{-6.63}$ &  $-$ & $-0.09^{+0.32}_{-0.33}$ & $0.98$ \\ [1ex]
\hline
GW170304 & $-$ & $42.9^{+9.79}_{-9.73}$ & $-$ & $31.6^{+7.5}_{-7.5}$ &  $-$ & $0.19^{+0.31}_{-0.31}$& $0.985$ \\ [1ex]
\hline
GW170817A & $-$ & $56^{+16}_{-10}$ & $-$ & $40^{+10}_{-11}$ &  $-$ & $0.5^{+0.2}_{-0.2}$ & $0.86$  \\ [1ex]
\hline
GW170425 & $-$ & $46.7^{+14.9}_{-14.92}$ & $-$ & $29.9^{+9.87}_{-9.91}$ &  $-$ & $0.05^{+0.42}_{-0.42}$ & $0.77$ \\ [1ex]
\hline 
GW151216 & $-$ & $32.3^{+9.7}_{-9.7}$ & $-$ & $20.47^{+5.76}_{-5.76}$ &  $-$ & $0.8^{+0.18}_{-0.18}$& $0.71$  \\ [1ex]
\hline
GW170202 & $-$ & $29.87^{+11.46}_{-11.45}$ & $-$ & $14.32^{+4.48}_{-4.48}$ &  $-$ & $-0.16^{+0.34}_{-0.34}$ & $0.68$ \\ [1ex]
\hline
GW170403 & $-$ & $45.5^{+10.25}_{-10.25}$ & $-$ & $32.7^{+8.29}_{-8.25}$ &  $-$ & $-0.63^{+0.39}_{-0.38}$ & $0.56$ \\ [1ex]
\hline
\hline
\end{tabular}
\caption{Parameters of the BBH mergers detected during LVC's O1 and O2 by LVC (left columns, \citet{LIGO_cat}) and IAS (right columns, \citet{IAS2,IAS3}). The parameters are median values with $90\%$ credible intervals.}
\label{tab:data}
\end{center}
\end{table*}

\section
{The Model Distribution for Field Binaries}
\label{App_model}
The model distribution, $p_{th}\brk*{\chie; \lambda}$, 

$\lambda = \brk[c]*{{\rm{SA}_{0,syn}/\rm{DA}_{0,syn}}, m_1, m_2,  t_{c,\min}, t_{w}, t_{\WR}}$, is derived under the assumptions given in the main text. 
\label{app1}
 
We take the BBH formation rate per volume element per unit comoving time, $R\brk*{z}$,   to follow the star formation rate (we also consider other rates, see Appendix \ref{Additional Tests}).

The mergers' rate  follows the formation rate with a delay $t_{c}$ whose probability is assumed to be $p_{obs}\brk*{t_c} \propto t_c^{-1}$ for $t_{c}>t_{c,\min}$. This  allows us to define the probability that the merger occurred at redshift $z_c$ as:
\begin{align}
    p\brk*{z_c} = \frac{1}{t_c\brk*{z_c}} \frac{R\brk*{z_c}}{1+z_c} \frac{\dd t_c}{\dd z_c}
\ .
\end{align}
We approximate, implicitly, that all mergers take place now (relaxing this assumption and assuming that the mergers take place between $z=0$ and $z=0.5$ doesn't change our results).

\begin{figure}[htb!]
   \centering
    \includegraphics[scale=0.4]{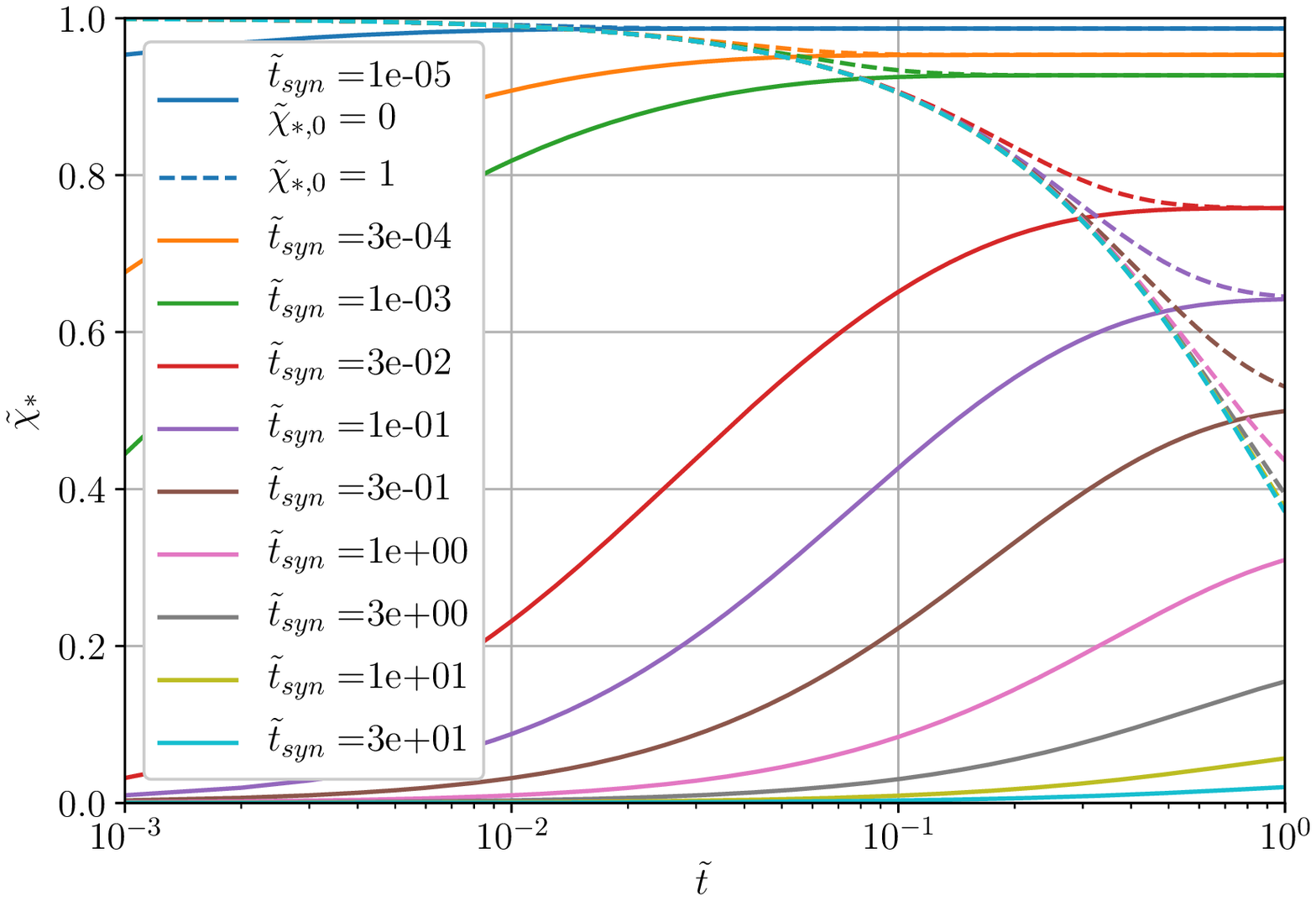}
    \caption{$\tilde \chi\equiv \chi_{*}/\chi{syn}$
as a function of $\tilde t \equiv t/t_w$ for different values of $\tilde t_{syn} \equiv \chi_{syn}^{-5/3} (t_{syn}/t_w)$.  }
    \label{fig:SM_evolution}
 \end{figure}

For a given $t_{c}$ and fixed $t_{w}, t_{\WR}$ we compute the final stellar spin, $\chi_{*}(t_{\WR})$ by integrating (Eq. 5 of the main text):
\begin{equation} 
\label{eq:chi_star_2}
	\frac{\textrm{d} {\chi_{*}}}{\textrm{d}{t}} = \frac {({\chi_{syn}}-  \chi_{*})^{{8}/{3}}}{{t}_{syn}(t_c)}
	 - \frac{{\chi}_{*}}{{t_w}} \ , 
\end{equation}
from 0 to $t_{\WR}$.
As initial conditions we take 
\begin{equation} 
\chi_*(0) = \begin{cases} 0 
&\mbox{unsynchronized } \ , \\ 
\chi_{syn} & \mbox{synchronized }   \ .
\end{cases}
\end{equation}
Fig. \ref{fig:SM_evolution} depicts the results of this integration in terms of $\chi_{*}/\chi_{syn}$,
as a function of $t/t_w$  for different ratios of $\tilde t_{syn} \equiv \chi_{syn}^{-5/3} (t_{syn}/t_w$). 
 
The BH spin after the collapse is then given by: 
\begin{equation} 
\chi_{_{BH}} = \min\brk[c]*{\chi_{*}(t_{\WR}), 1} \ .
\label{eq:BH_spin} 
\end{equation}

Under the assumption that $\chi_{_{BH}}$ is deterministic w.r.t it's parameters, we may write it's distribution using the chain rule:
\begin{equation}
    p\brk*{\chi_{_{BH}}; \lambda} = \frac{1}{t_c\brk*{z_c}} \frac{R\brk*{z_c}}{1+z_c}
    \frac{\dd \chi_{_{BH}}}{\dd t_c} \frac{\dd t_c}{\dd z_c},
\label{eq:p} 
\end{equation}
where we calculate numerically the derivative
${\dd \chi_{_{BH}}}/{\dd t_c}$, following the integration of Eq. \ref{eq:chi_star} above.
 
To obtain the final $\chie$ distribution:  
\begin{equation}
    \chie= \frac{\chi_{_{BH,1}} + q \chi_{_{BH,2}}}{\brk*{1+q}} \ , 
\label{eq:chie} 
\end{equation}
we  consider the different scenarios separately. 
\begin{itemize}
    \item {\rm{SA}}: $\chi_{_{BH,1}}$ is distributed as  {\it flat} isotropic  and $\chi_{_{BH,2}}$ is given by equation \eqref{eq:p}. To find the resulting distribution of $\chie$ we sample each (from the respective distribution) and calculate the empirical distribution of their weighted sum.
    \item {\rm{DA}}: Using the above procedure, for a given $t_c$ ,we calculate $\chi_{_{BH,1}}$ and $\chi_{_{BH,2}}$.  Using the numerical values of the derivative, ${\dd \chi_{eff}}/{\dd t_c}$, we obtain the distribution: 
\begin{equation}
p\brk*{\chi_{eff}; \lambda} = \frac{1}{t_c\brk*{z_c}} \frac{R\brk*{z_c}}{1+z_c}
    \frac{\dd \chi_{eff}}{\dd t_c} \frac{\dd t_c}{\dd z_c} \ .
\label{eq:p_chi} 
\end{equation}
\end{itemize}
        \setcounter{figure}{0}
        \renewcommand{\thefigure}{B\arabic{figure}}%

\section{Additional Tests}
\label{Additional Tests}
 
\begin{figure}[htb!]
   \centering
    \includegraphics[scale = 0.8]{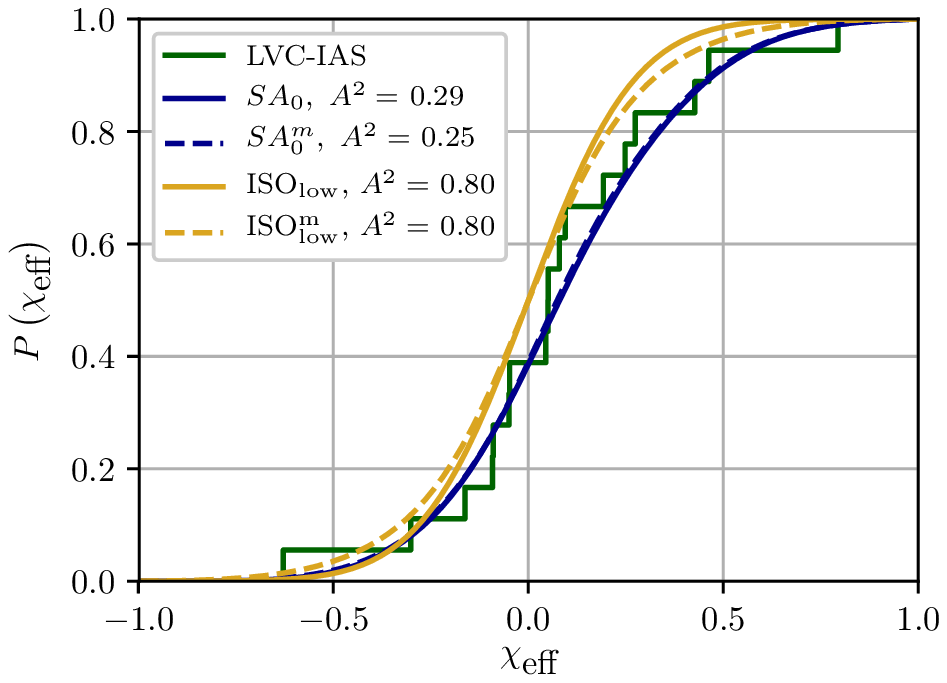}
    \caption{The models evaluated using a mixture of mass ratios, $m_1, m_2$. That is for each model we find it's prediction to each of the $N=18$ observed masses and consider the average of this predictions as the final probability of the model. We denote the mixture probabilities a superscript $m$. }
    \label{fig_sup:mass_dep}
 \end{figure}

{\it The Mass Distribution:}
The masses used in the estimates are   the average  values of the sample: $\overline {m}_1=38M_\odot$ and $\overline{m}_{2}=24M_\odot$. To explore the effect of the different masses, we also use the masses of the observed events and sample over the mass distribution. The results are shown in Fig. \ref{fig_sup:mass_dep} for the SA distribution  and  for the isotropic models whose $\chie$ distribution is affected (becomes broader) when mass ratio is taken into account. 
We find that the results are almost the same as those obtained using the average mass and  mass ratio.  

\begin{figure}[htb!]
   \centering
    \includegraphics[scale = 0.8]{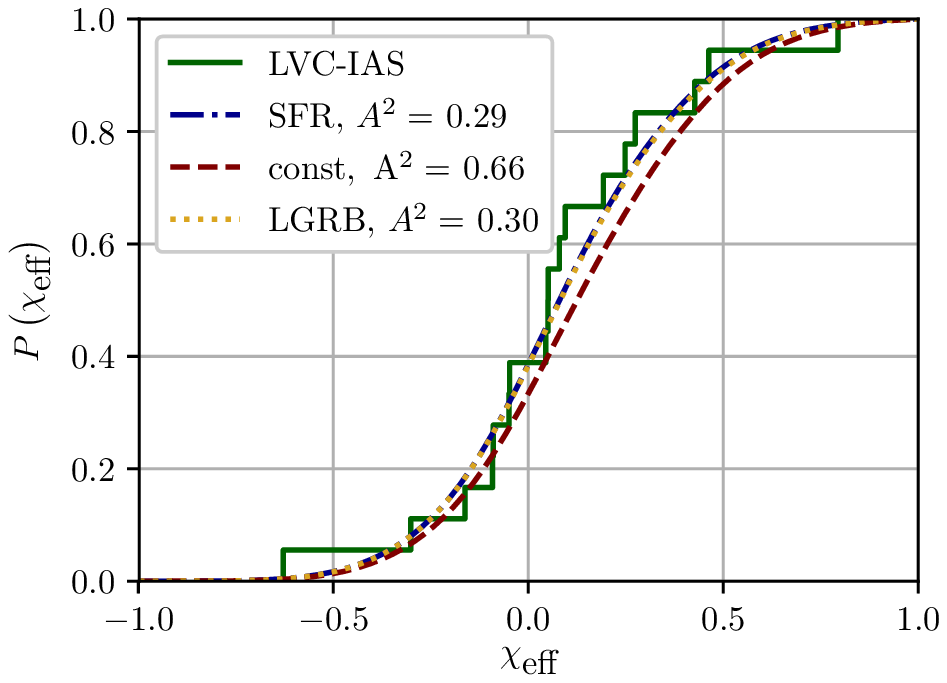}
    \caption{The effect of the rate function, $R\brk*{z}$ on the model prediction. All rates are evaluated under the $\rm{SA}_{0}$ scenario.}
    \label{fig_sup:rate}
 \end{figure}

 {\it The Event Rate:} We use  the star formation rate as the event rate for the formation of BBH. We also consider the possibility that BBH follow the long GRB (LGRB) rate, as it was suggested that long GRBs indicate the formation of a BBH \citep{KT_ANL}, and a (ad hoc) constant formation rate.
 Fig. \ref{fig_sup:rate} demonstrates that  the resulting distribution is practically independent of the assumption on the SFR. 
 
\begin{figure}[htb!]
   \centering
    \includegraphics[scale = 0.8]{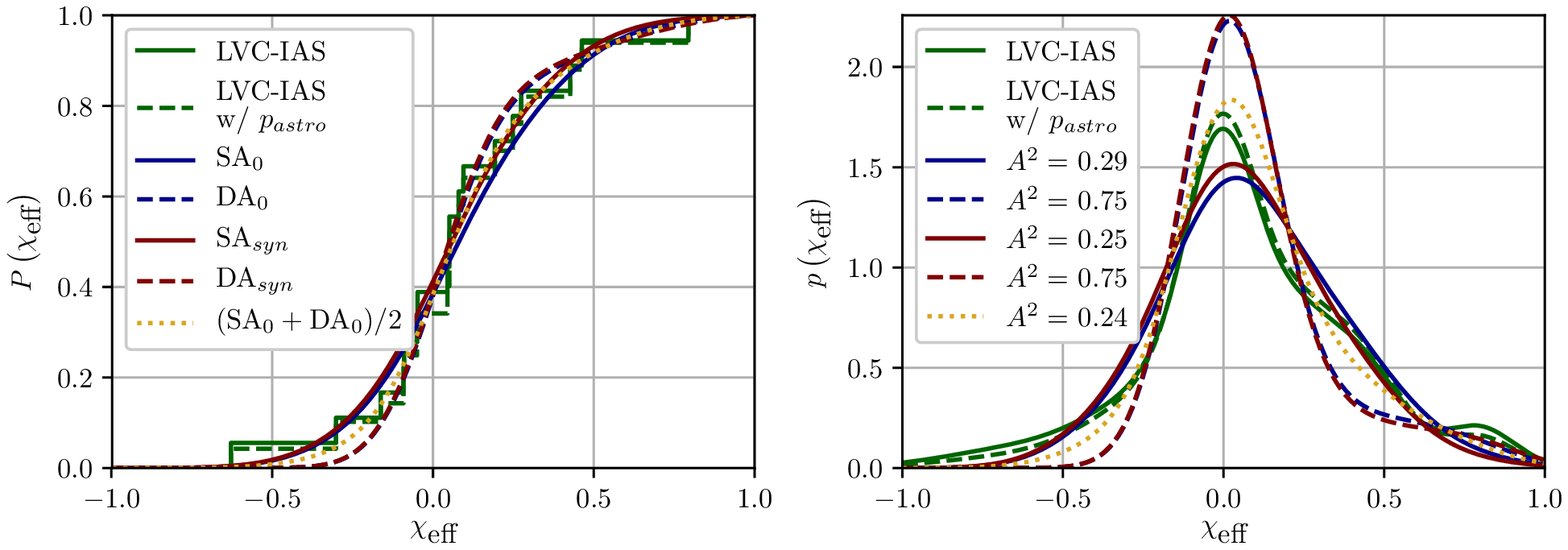}
    \caption{{ The effect of weighting the observed events by $p_{astro}$ on the obtained observed distribution and goodness of fit to the models (calculated with the same parameters).} }
    \label{fig_sup:p_astro}
 \end{figure}

 { {\it $p_{astro}$ weight:} The observed events are given a probability that the event is of astrophysical, $p_{astro}$. Weighting the events using this value to obtain the observed distribution does not affect the goodness of fit ($A^2$ score)  of our models, see Fig. \ref{fig_sup:p_astro}.} 

\clearpage

\end{document}